\documentclass[10pt,a4paper,twoside]{article}
\usepackage{epsfig}
\usepackage{baltlat5}
\usepackage{wrapfig}
\begin{document}
\ \
\vspace{-2.5mm}

\setcounter{page}{511}

\titlehead{Baltic Astronomy, vol.\ts 14, 511--525, 2005.}

\titleb{RADIAL VELOCITIES OF POPULATION II BINARY STARS. II.}

\begin{authorl}
\authorb{A. Bartkevi\v cius}{1,3} and
\authorb{J. Sperauskas}{1,2}
\end{authorl}

\begin{addressl}
\addressb{1}{Institute of Theoretical Physics and Astronomy,
Vilnius University,\\ Go\v{s}tauto 12, Vilnius, LT-01108, Lithuania}

\addressb{2}{Vilnius University Observatory, \v Ciurlionio~29, Vilnius,
 LT-03100,\\ Lithuania}

\addressb{3}{Department of Theoretical Physics, Vilnius Pedagogical University,
\\ Student\c u~39, Vilnius, LT-08106, Lithuania}

\end{addressl}

\submitb{Received 2005 November 21}

\begin{summary} Here we publish the second list of radial velocities for
91 {\it Hipparcos} stars, mostly high transverse velocity binaries
without previous radial velocity measurements.  The measurements of
radial velocities are done with a CORAVEL-type radial velocity
spectrometer with an accuracy better than 1 km/s.  We also present the
information on eight new radial velocity variables -- HD 29696,
HD 117466\,AB, BD +28 4035\,AB, BD +30 2129\,A, BD +39 1828\,AB,
BD +69 230\,A, BD +82 565\,A and TYC 2267-1300-1 -- found from our
measurements.  Two stars (HD\,27961\,AB and HD\,75632\,AB) are suspected
as possible radial velocity variables.  \end{summary}


\resthead{Radial velocities of population II binary stars. II}{A.
Bartkevi\v cius, J. Sperauskas}

\sectionb{1}{INTRODUCTION}

We started radial velocity measurements of population II (thick disk +
halo) binary stars in 1988 (Bartkevi\v cius \& Sperauskas 1990, 1994),
as a part of the general program of investigation of population II stars
selected from MDSP (Bartkevi\v cius 1980, 1984), MDPH (Bartkevi\v cius
1992), POP2 (Bartkevi\v cius \& Bartkevi\v cien\. e 1993) and other
Population II star catalogs (Bartkevi\v cius 1994).  Radial velocity
measurements for 245 single stars and double or multiple star components
were published by Bartkevi\v cius et al.  (1992) and Bartkevi\v cius \&
Sperauskas(1999).  A discussion of measurements of metal-deficient
binary star components and 15 suspected radial velocity variables was
presented in Bartkevi\v cius \& Sperauskas (1990).  Further discussion
of 33 suspected radial velocity variables is given by
Bartkevi\v cius \& Sperauskas (1994).  All the measurements have been
done by J. Sperauskas with the CORAVEL-type photoelectric radial
velocity speedometer built by Tokovinin (1987) at the Sternberg
Astronomical Institute in Moscow and attached to the 1 meter reflector
of the Institute of Theoretical Physics \& Astronomy, Vilnius, placed at
the Maidanak Observatory in Uzbekistan.

In 1998, when J. Sperauskas constructed own CORAVEL-type radial
velocity spectrometer, we started a new program of radial velocity
measurements devoted almost completely to the population II double and
multiple stars.  Mostly \hbox{{\it Hipparcos}} high transverse velocity
stars without radial velocity measurements were observed.  In 2002 we
published the first list of radial velocities of 114 stars (Sperauskas
\& Bartkevi\v cius 2002, Paper I).  In this publication we present
radial velocities for the second list of 91 stars.

\sectionb{2}{SELECTION OF PROGRAM STARS}

Principles of the selection of stars for radial velocity observations
were practically the same as in Paper I. For the selection of the
program stars we used the {\it Hipparcos} Catalogue Double Star Annexes
(ESA 1997), taking the stars with transverse velocity greater than
60 km/s.  Some suspected {\it Hipparcos} double stars (Solution S) were
also observed.  In some cases for the isolation of high velocity
subdwarf binaries, the reduced proper motion diagram $H_{V}$, $B$--$V$
(Luyten 1922; Jones 1972; Chiu 1980) was used.  Several high transverse
velocity thick disk orbital binaries (Bartkevi\v cius \& Gudas 2001,
2002) were also included.  Only stars without radial velocity
measurements or with the velocities of low quality, as well as
suspected radial velocity variables were included in the program.  To
check the availability of radial velocities the following sources were
consulted:  the {\it General Catalog of Averaged Stellar Radial
Velocities for Galactic Stars} by Barbier-Brossat \& Figon (2000), the
{\it Bibliographic Catalogue of Stellar Radial Velocities (1991--1994)}
by Malaroda et al.  (2000) and the {\it Bibliographic Catalogue of
Stellar Radial Velocities for Population II and Late-type Stars in the
Galaxy} by Bartkevi\v cius (2000).

\sectionb{3}{RADIAL VELOCITY MEASUREMENTS AND DATA REDUCTION}

As in Paper I, radial velocity measurements were made with the
CORAVEL-type radial velocity spectrometer constructed by J. Sperauskas
at the Vilnius University Observatory and based on the principles
developed by Griffin (1967).  The spectrometer is described by Upgren et
al.  (2002).  91 selected program stars were observed by J. Sperauskas
during four observing runs.  During the first (1998 July 16--27, JD
2451010--2451022) and the second runs (2000 August 22 -- September 8, JD
2451779--2451795) observations were carried out with the 1.65 m
telescope of the Mol{\. e}tai Observatory (Lithuania).  During the third
run (2000 December 13--16, JD 2451892--2451894) a 1.5 m RTT telescope of
the Tubitak National Observatory (Turkey) was used.  During the fourth
run (2003 October 27 -- December 28 , JD 2452940--2453002) the 1.5, 1.53
and 2.3 m reflectors of the Steward Observatory (Arizona) were used.

The zero point and time-dependent drift of the spectrometer each night
were determined by observing several IAU radial velocity standard
stars from the Udry et al.  (1999) list.  Weighted mean radial
velocities $<{V_r}>$, their standard error $\varepsilon$, the internal
and the external errors $I$ and $E$ were calculated according to the
precepts by Jasniewicz \& Mayor (1988).

\sectionb{4}{ RESULTS AND DATA ANALYSIS}

Table 1 lists the basic data for the program stars:  the running number,
the star name, the identification in CCDM or WDS catalogs, the
identification in the {\it Hipparcos} catalog, equatorial coordinates,
Johnson's $V$ magnitude and color-index $B$--$V$, and the spectral type.
Notes to Table 1 contain some additional information:

\begin{center}
\vbox{\scriptsize
\tabcolsep 1pt
\begin{tabular}{rlllccrll}
\multicolumn{9}{c}{\parbox{125mm}{\baselineskip=8.5pt
{\smallbf Table 1.}{ \small Basic data of the program stars.}}}\\
\noalign{\vskip1mm}
\tablerule
No. &  ~~Star name  &  CCDM/WDS  &  HIP & RA (2000) & DEC (2000) & $V$~
& $B$--$V$   &  ~~~Sp \\
    &             &            &      & h~~~m~~~s &
$\circ$~~~$\prime$~~~$\prime\prime$  &  mag  &  mag  &    \\
\tablerule
\noalign{\vskip1mm}
 1. &    HD\,225220\,AB  &CCDM 00046+3416AB  &   374  & 00 04 40.1 & +34 15 54 & 7.02  & 1.10  & G8\,III             \\
 2. &   HIP\,375    &CCDM 00046+3416C   &   375  & 00 04 40.2 & +34 16 17 & 9.08  &       &                   \\
 3. &   HDE\,236325    &                   &        & 00 10 17.9 & +56 34 12 & 8.43  & 1.37  & K5, K\,sl\,wl       \\
 4. &    BD\,+38   35\,AB  &                   &  1737  & 00 21 58.6 & +39 38 17 &10.28  & 0.61  &                   \\
 5. &   HDE\,232169    &                   &        & 00 22 48.9 & +53 11 33 & 8.91  & 0.64  & G0, G\,mod\,wl      \\
\end{tabular}
}
\end{center}

\begin{center}
\vbox{\scriptsize
\tabcolsep 1pt
\begin{tabular}{rlllccrll}
\multicolumn{9}{c}{\parbox{125mm}{\baselineskip=8.5pt
{\smallbf Table 1.}{ \small \hskip 1.5mm Continued}}}\\
\noalign{\vskip1mm}
\tablerule
No. &   ~~Star name  &  CCDM/WDS  &  HIP & RA (2000) & DEC (2000) & $V$~ &
$B$--$V$   &  ~~~Sp \\
    &             &            &      & h~~~m~~~s &
$\circ$~~~$\prime$~~~$\prime\prime$  &  mag  &  mag  &    \\
\tablerule
\noalign{\vskip1mm}
64. &    HD\,74861\,AB  &CCDM 08462-1422AB  & 43038  & 08 46 09.1 & --14 22 05& 8.35  & 0.51  & F8\,V              \\
65. &    BD\,+46 1436    & WDS 08479+4532AB  & 43189  & 08 47 55.3 & +45 31 39 & 9.22  & 0.77  &                   \\
66. &    HD\,75632\,AB  &CCDM 08555+7048AB  & 43820  & 08 55 24.8 & +70 47 39 & 8.05  & 1.39  & K5\,V\,comp         \\
67. &    BD\,+49 2588    &CCDM 17087+4945A   & 83885  & 17 08 41.8 & +49 45 19 & 9.32  & 0.65  & G0                \\
68. &    BD\,+68  986\,A   &CCDM 18161+6839A   & 89523  & 18 16 04.0 & +68 38 55 &10.13  & 0.69  & G5                \\
\end{tabular}
}
\end{center}

\begin{center}
\vbox{\scriptsize
\tabcolsep 1pt
\begin{tabular}{rlllccrll}
\multicolumn{9}{c}{\parbox{125mm}{\baselineskip=8pt
{\smallbf Notes to Table 1.}{}}}\\
\noalign{\vskip1mm}
\tablerule
No. &  ~~Star name  &  Notes  \\
\tablerule
\noalign{\vskip2mm}
 1.&    HD    225220\,AB  & HIP-C, AB: $PA$ = 143.4$\degr$, $d$ = 0.724\arcsec, $\Delta Hp$ = 0.284 mag;  D: GSC 2267-1300,\\
   &                    & AD: $PA$ = 238$\degr$, $d$ = 95.7\arcsec, $\Delta m$ = 1.7 mag, Var Vr \\
 2.&   HIP       375    & HIP-C, AC: $PA$ = 1.9$\degr$, $d$ = 22.700\arcsec;  AB-C: optical     \\
 3.&   HDE    236325    & BD +55 13,  SAO 21158, TYC 3660-1172-1;
discussion: BS 2003       \\
 4.&    BD  +38   35\,AB  & HIP-X                                                                                           \\
 5.&   HDE    232169    & BD +52 55, SAO 21335, TYC 3653-1448-1;  suspected in variability: \\
\end{tabular}
}
\end{center}

\begin{center}
\vbox{\scriptsize
\tabcolsep 1pt
\begin{tabular}{rlllccrll}
\multicolumn{9}{c}{\parbox{125mm}{\baselineskip=8pt
{\smallbf Notes to Table 1.}{\small \hskip 1.5mm Continued}}}\\
\noalign{\vskip1mm}
\tablerule
No. &  ~~Star name  &  Notes  \\
\tablerule
\noalign{\vskip2mm}
26.&    HD     20289\,AB  & HIP-C, AB: $PA$ = 331.7$\degr$, $d$ = 0.918\arcsec, $\Delta m$ = 1.55 mag                                                \\
27.&    HD     20369\,AB  & G 77-37, LTT 11077; HIP-C, AB: $PA$ = 270.0$\degr$, $d$ = 2.507\arcsec, $\Delta m$ = 2.41 mag                           \\
28.&    BD  +12  472\,AB  & HIP-S            \\
29.&   HIP     16332\,AB  & HIP-C, AB: $PA$ = 259.0$\degr$, $d$ = 0.249\arcsec, $\Delta Hp$ = 0.14 mag                                              \\
30.&   HDE    232816    & BD +52 705, SAO 24134, TYC 3716-490-1; discussion: BS 2003       \\
31.&    BD  +16  492\,AB  & HIP-C, AB: $PA$ = 289$\degr$, $d$ = 2.234\arcsec, $\Delta m$ = 3.49 mag                                                 \\
\end{tabular}
}
\end{center}

\begin{center}
\vbox{\scriptsize
\tabcolsep 1pt
\begin{tabular}{rlllccrll}
\multicolumn{9}{c}{\parbox{125mm}{\baselineskip=8pt
{\smallbf Notes to Table 1.}{\small \hskip 1.5mm Continued}}}\\
\noalign{\vskip1mm}
\tablerule
No. &  ~~Star name  &  Notes  \\
\tablerule
\noalign{\vskip2mm}
81.&   HDE    235634    & BD +43 2713, SAO 33787, TYC 3968 2435 1;  MD-F7\,V, $(Y-V)_0$ = 0.49 mag,\\
   &                    & $E_{Y-V}$ = 0.03 mag,  ${M}_V$ = 4.0 mag, [Fe/H] = --0.24 [Viln];  discussion: BS 2003               \\
82.&    BD  +14 4697\,AB  & G 126-42; HIP-C, AB: $PA$ = 155.1$\degr$, $d$ = 1.243\arcsec, $\Delta Hp$ = 0.03 mag                                    \\
83.&   HDE    239892\,AB  & BD +59 2488, SAO 34178, TYC 4263-1795-1, ADS 15478;  unresolved duplicity \\
   &                    & [TYC]; AB: $PA$ = 272$\degr$, $d$ = 0.7\arcsec, $\Delta m$ = 0.0 mag; MD-F7\,V, $(Y-V)_0$ = 0.49\\
   &                    & mag,  $E_{Y-V}$  = 0.01 mag,  ${M}_V$ = 3.7 mag, [Fe/H] = --0.37  [Viln]; discussion: \\
\end{tabular}
}
\end{center}

\vskip3mm

\begin{center}
\vbox{\scriptsize
\tabcolsep 5pt
\begin{tabular}{rlrlccrll}
\multicolumn{8}{c}{\parbox{90mm}{\baselineskip=8.5pt
{\smallbf Table 2.}{ \small Mean radial velocities.}}}\\
\noalign{\vskip1mm}
\tablerule
No. &  ~~Star name  & $<{V_r}>$  &  $e$ & $E$ & $E/I$ & $N$
& Notes  \\
    &               &  km/s      &  km/s & km/s &    &    & \\
\tablerule
\noalign{\vskip1mm}
 1. &    HD\,225220\,AB  &  --24.4 &    0.4  &   0.1  &   0.1  &     2 &             \\     [0.8pt]
 2. &   HIP\,375    &   --9.3 &    0.4  &   0.6  &   0.9  &     3 &             \\          [0.8pt]
 3. &   HDE\,236325    &   40.6 &    0.4  &   0.1  &   0.2  &     2 & BS 2003          \\        [0.8pt]
 4. &    BD\,+38 35\,AB  &   --9.4 &    0.6  &   0.4  &   0.5  &     2 &             \\     [0.8pt]
 5. &   HDE\,232169    &   --4.0 &    0.5  &   0.5  &   0.7  &     2 & BS 2003          \\
\end{tabular}
}
\end{center}
\newpage

\begin{center}
\vbox{\scriptsize
\tabcolsep 5pt
\begin{tabular}{rlrlccrll}
\multicolumn{8}{c}{\parbox{90mm}{\baselineskip=8.5pt
{\smallbf Table 2.}{ \small \hskip 1.5mm Continued}}}\\
\noalign{\vskip1mm}
\tablerule
No. &  ~~Star name  & $<{V_r}>$  &  $e$ & $E$ & $E/I$ & $N$
& Notes  \\
    &               &  km/s      &  km/s & km/s &    &    & \\
\tablerule
\noalign{\vskip1mm}
35. &   HIP\,17876    &  --10.3 &    0.8  &        &        &     1 &             \\        [0.8pt]
36. &    HD\,26735\,A   &   64.0 &    0.5  &   0.9  &   1.3  &     3 &             \\       [0.8pt]
37. &    HD\,26735\,B   &   63.7 &    0.3  &   0.6  &   1.0  &     4 &             \\         [+0.8pt]
38. &    HD\,27961\,AB  &   36.3 &    0.7  &        &        &     1 &             \\         [+0.8pt]
39. &    BD\,+03  609    &   53.0 &    0.5  &   0.6  &   0.9  &     2 &             \\
\end{tabular}
}
\end{center}

\begin{center}
\vbox{\scriptsize
\tabcolsep 2pt
\begin{tabular}{rlcrc|rlcrc}
\multicolumn{10}{c}{\parbox{120mm}{\baselineskip=8.5pt
{\smallbf Table 3.}{ \small Individual radial velocities.}}}\\
\noalign{\vskip1mm}
\tablerule
No. &  ~~Star name  &  HJD  &  ${V_r}$~~ & ${\varepsilon}1 $ &
No. &  ~~Star name  &  HJD  &  ${V_r}$~~ & ${\varepsilon}1 $  \\
    &               & +2400000 &   km/s   &  km/s           &
    &               & +2400000 &   km/s   &  km/s           \\
\tablerule
\noalign{\vskip1mm}
 1. &  HD\,225220\,AB  &  51794.481  &  --24.3  &  0.6    &   ~~34.    &                &  52994.830  &   --6.9  &    0.6                \\
   &                &  52950.633  &  --24.4  &    0.6    &    ~~35. & HIP\,17876    &  52992.880  &  --10.3  &    0.8                \\
 2. & HIP\,375    &  51783.569  &   --9.8  &    0.7       &    ~~36. &  HD\,26735\,A   &  51893.445  &   64.3  &    0.7                \\
   &                &  51794.486  &   --8.6  &    0.7    &       &                &  51893.453  &   64.8  &    0.8                \\
   &                &  52950.672  &   --9.4  &    0.6    &       &                &  52991.865  &   63.0  &    0.7                \\
\end{tabular}
}
\end{center}

\newpage

\begin{center}
\vbox{\scriptsize
\tabcolsep 2pt
\begin{tabular}{rlcrc|rlcrc}
\multicolumn{10}{c}{\parbox{125mm}{\baselineskip=8.5pt
{\smallbf Table 3.}{ \small \hskip 1.5mm Continued}}}\\
\noalign{\vskip1mm}
\tablerule
No. &  ~~Star name  &  HJD  &  ${V_r}$~~ & ${\varepsilon}1 $ &
No. &  ~~Star name  &  HJD  &  ${V_r}$~~ & ${\varepsilon}1 $  \\
    &               & +2400000 &   km/s   &  km/s           &
    &               & +2400000 &   km/s   &  km/s           \\
\tablerule
\noalign{\vskip1mm}
68. &  BD\,+68  986\,A   &  51784.338  &    1.5  &    0.8   &  ~~81.    &                &  51022.427  &   16.6  &    0.7                \\
   &                &  51794.337  &    2.0  &    0.7       &      &                &  51784.436  &   17.3  &    0.7                \\
69. &  BD\,+75  641    &  51016.382  & --118.8  &    0.7    &      &                &  51795.384  &   18.0  &    0.7                \\
   &                &  51018.403  & --119.4  &    0.7      &   ~~82. &  BD\,+14 4697\,AB  & 51779.467   & --57.0   &   0.7             \\
   &                &  51018.423  & --119.3  &    0.7      &      &                & 51786.446   & --55.6   &   0.7                 \\
\end{tabular}
}
\end{center}


\hbox{
\vtop{\hsize61mm
\centerline{\psfig{figure=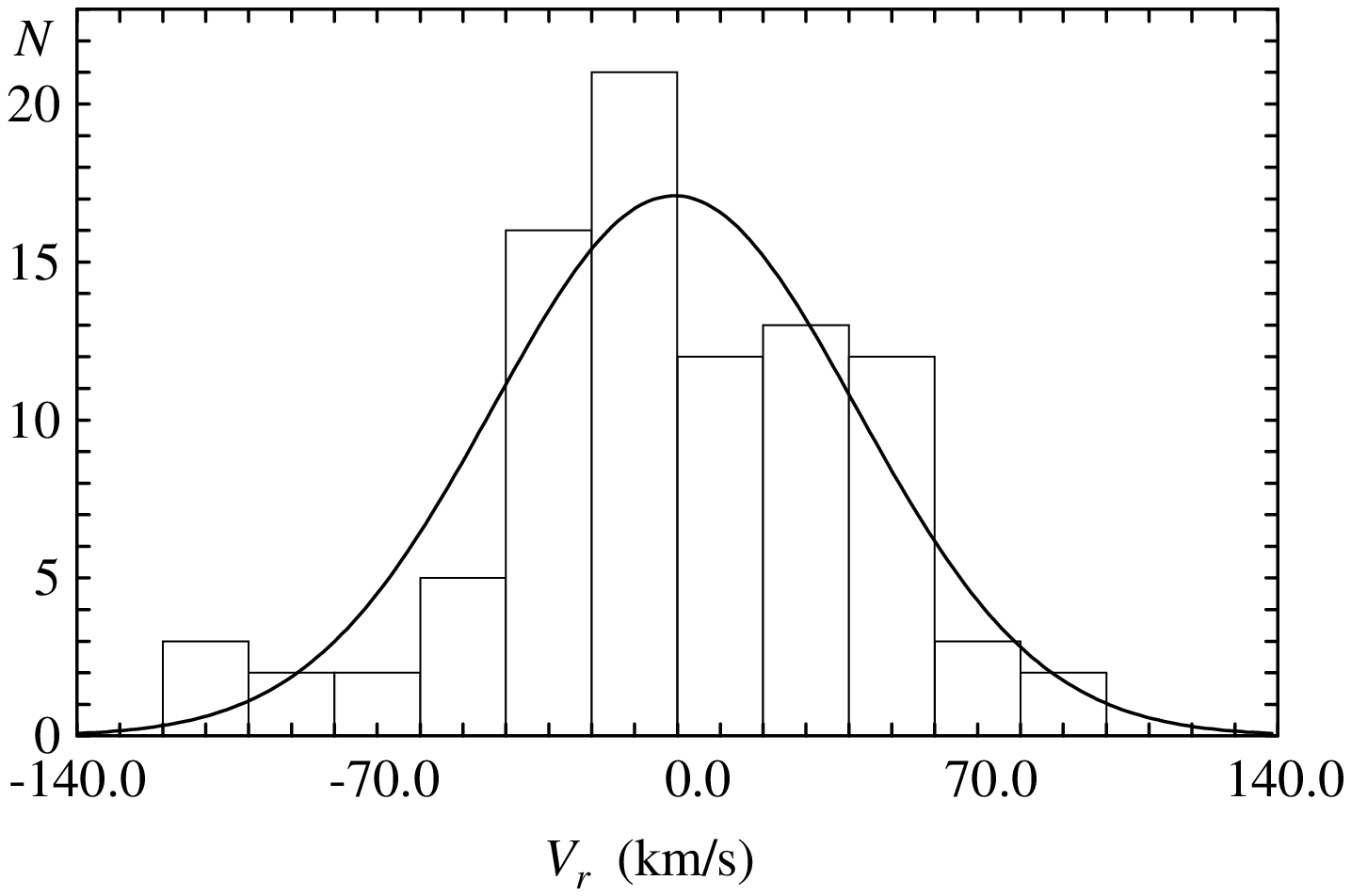,width=61mm,angle=0,clip=}}
\vspace{1mm}
\captionr{1}{The histogram  of the mean radial velocity values
$<{V_r}>$.}
}
\hspace{.2mm}
\vtop{\hsize61mm
\centerline{\psfig{figure=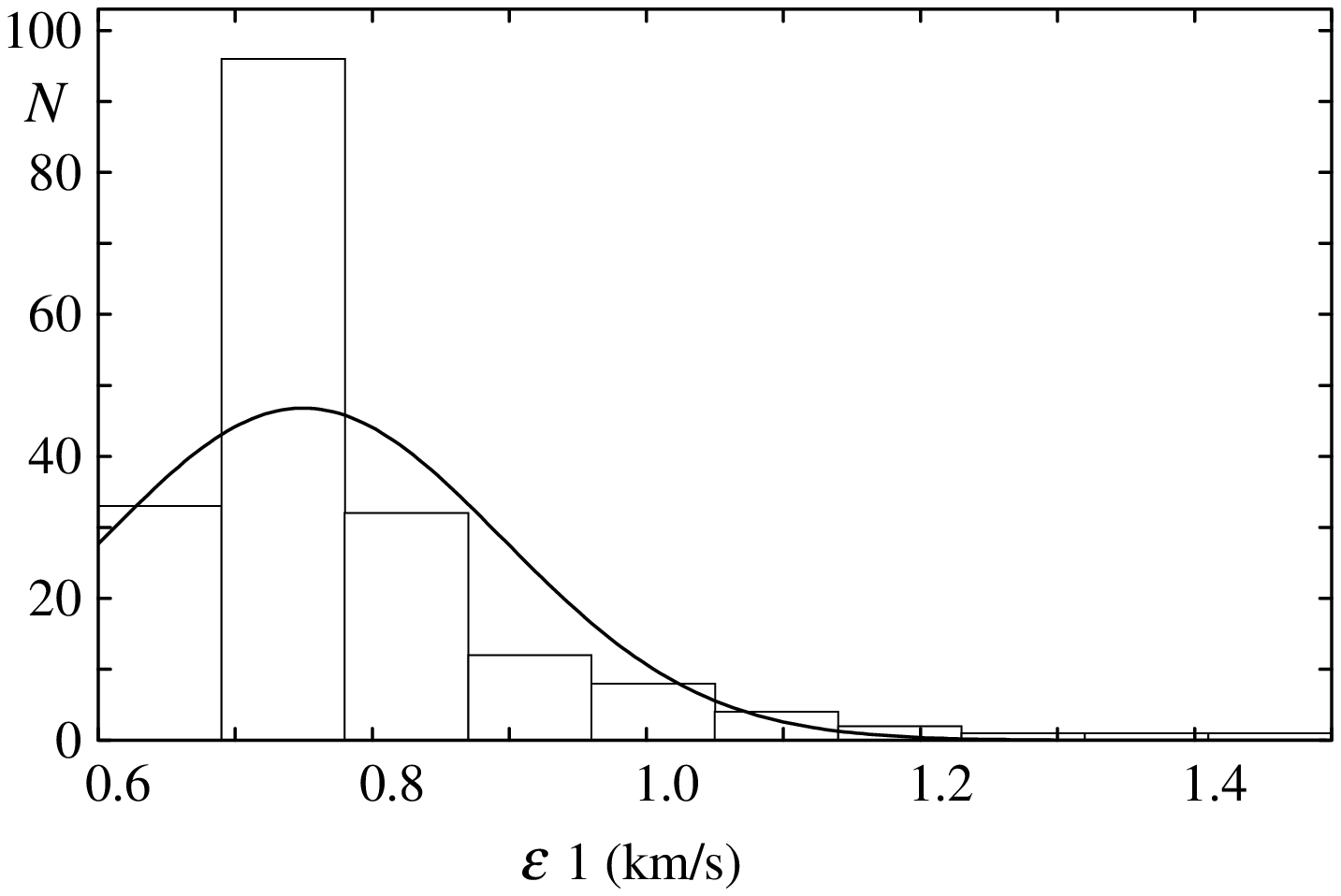,width=61mm,angle=0,clip=}}
\vspace{1mm}
\captionr{2}{The histogram  of the individual radial velocity
measurement error $\varepsilon$1.}
}
}
\vskip2mm

\noindent the type of {\it
Hipparcos} binary solutions, the position angle of visual binary, the
angular distance between the components and the magnitude differences.

Mean radial velocities $<{V_r}>$, their standard error $\varepsilon$,
external error $E$, external-to-internal error ratio $E/I$ and the
number of measurements $n$ are presented in Table 2. For the sake of
completeness the results for 11 Bidelman's weak-lined stars (Bidelman
1998; Loth \& Bidelman 1998), discussed in Bartkevi\v cius \& Sperauskas
(2003, hereafter BS 2003), are included.

Individual values of radial velocities $V_r$, together with their errors
$\varepsilon$1 and the Julian Day date of observations, are given in
Table 3.

A histogram of the mean radial velocity values is shown in Figure 1.
Despite the fact that the selected program stars have high transverse
velocities, the range of velocities is not very broad: the
maximal radial velocity is +99.1 km/s, the minimal --118.6 km/s.

\begin{wrapfigure}[19]{r}[0pt]{62mm}
\centerline{\psfig{figure=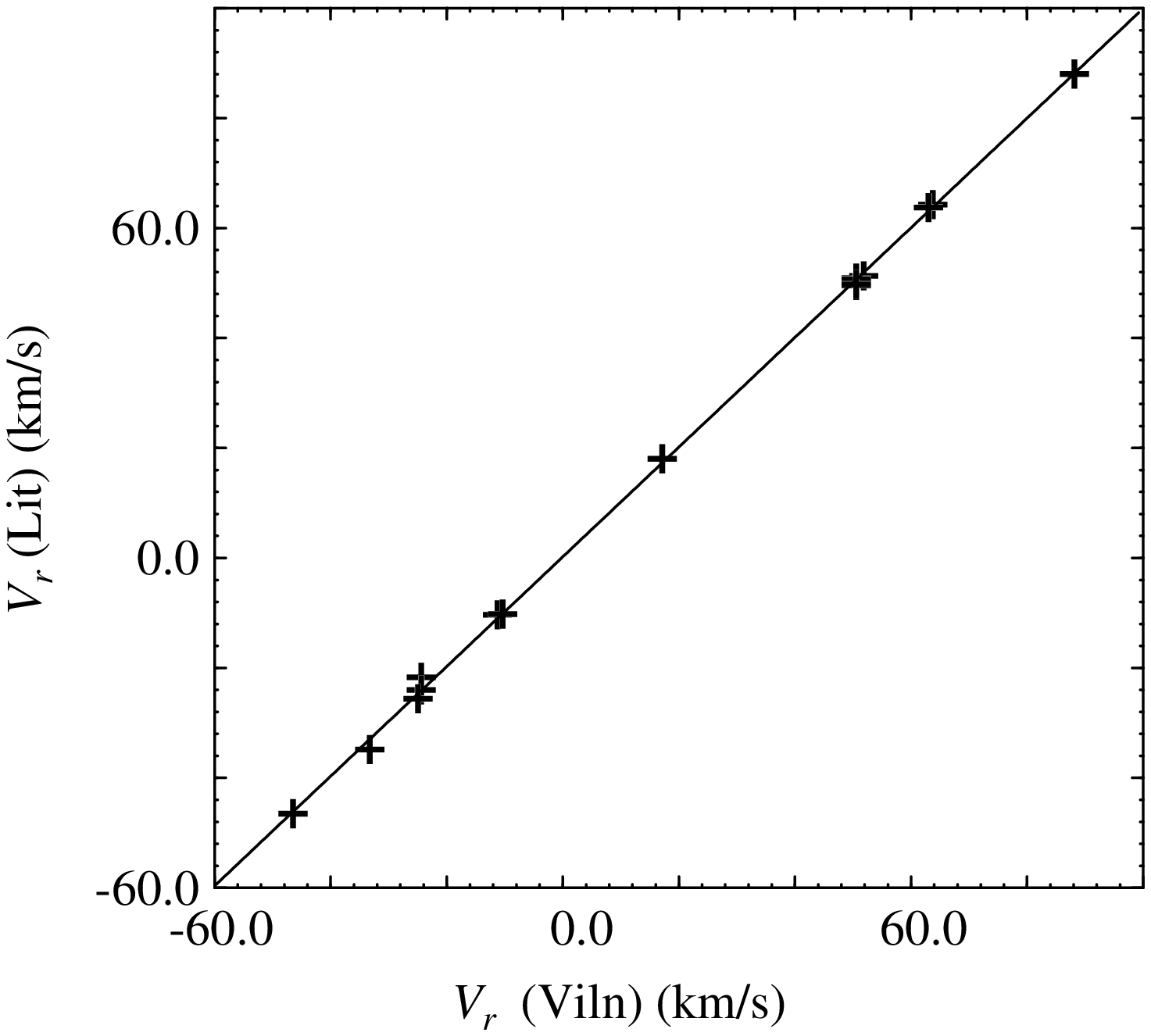,width=61truemm,angle=0,clip=}}
\vskip1mm
\captionr{3}{Comparison of our radial velocities ${V_r}$ (Viln) and the
literature values ${V_r}$ (Lit).}
\end{wrapfigure}

In Figure 2 we show the distribution of the individual radial velocity
measurement errors $\varepsilon$1.  The mean value for 190 measurements
is $<{\varepsilon}1>$ = 0.75$\pm$0.01 km/s.  The errors $\varepsilon$1
do not depend much on the observed star magnitude -- for faintest
observed stars they are larger no more than 0.1--0.2 km/s.  The mean
error of the weighted mean velocities for 91 measured star (for stars
measured once we adopt $\varepsilon$ = $\varepsilon$1) is $<{
\varepsilon}>$ = 0.60$\pm$0.02 km/s.  The mean external error $E$ for 68
stars having the number of measurements $n$ $\ge$ 2 is practically the
same:  $<{E}>$ = 0.58$\pm$0.06 km/s.  The external-to-internal error
ratio is $<{E/I}>$=0.79$\pm$0.07.

The search in CDS, ADS, the archives of astrophysics e-prints, the
above-mentioned radial velocity catalogs and other literature sources
revealed 18 program stars with radial velocities measured earlier (Table
4). 17 of them either have the mean values $<{V_r}>$ with $\varepsilon$
$\le$ 2.7 km/s or are non-variable stars.  For these stars the mean
difference between our measurements and the literature values is quite
small:  $<\Delta Vr>$ = --0.17$\pm$0.23 km/s.  In Figure 3 our radial
velocities are compared with the literature values.

\sectionb{5}{RADIAL VELOCITY VARIABLES AND VISUAL DOUBLE STARS}

Analyzing radial velocities of visual binary stars, we found nine radial
velocity variables listed in Table 5. The given information is similar
to that presented in Table 1 for constant velocity stars.  Radial
velocity data for these stars are given in Table 6. Here we give the
same information as in Table 2, adding the radial velocity observation
time interval between the first and the last observation and the
difference between the maximum and the minimum values of the observed
radial velocities $<\Delta Vr>$.

The star BD +82 565A is a short period ($P$ = 12.7 d) single-lined
spectroscopic binary of the thick disk.  Spectroscopic orbit and
detailed discussion on it will be published in Sperauskas \& Bartkevi\v
cius (2005).  A preliminary spectroscopic orbit ($P$ = 5.3 d) is
determined for the B component of the visual binary BD +39 1828.

The stars BD +30 2129A and HD 117466\,AB (about their radial velocity
variability see in Paper I), TYC 2267-1300-1 and BD +28 4035\,AB
undoubtedly show variable radial velocities, but for the determination
of their orbits we need additional radial velocity observations.


\begin{center}
\vbox{\footnotesize
\begin{tabular}{rlccrl}
\multicolumn{6}{c}{\parbox{80mm}{\baselineskip=8pt
{\normbf\ \ Table 4.}{\norm\ Radial velocities from the literature.}}}\\
\tablerule
No. & Star name & ${V_r}$  & $\varepsilon$1 & $n$ & Reference \\
    &           & km/s     &  km/s          &     &           \\
\tablerule
\noalign{\vskip2mm}
 1.  & HD    225220\,AB   &--21.7 &2.7  & 3 & Woolley et al. (1981)
\\
    &                  &--23.98&0.29 &   &Famaey et al. (2005)               \\
 8.  &HDE    236523\,AB   &  0.  &4.3  & 4 &Duflot et al. (1992)              \\
16.  & HD     14106     &+51.36&0.23 &12 &Latham et al. (2002)               \\
17.  & HD     14202     &--10.3 &0.2  & 2 &Nordstrom et al. (2004)            \\
18.  &HIP     10774\,B    &--10.2 &0.2  & 2 &Nordstrom et al. (2004)            \\
24.  & BD  +00  549\,A    &+88.08&0.18 &20 &Latham et al. (2002)                  \\
    &                  &+83.8 &4.9  & 2 &Fouts, Sandage (1986)             \\
    &                  &+84   &7    & 1 &Ryan, Norris (1991)               \\
26.  & HD     20289\,AB   &+18.08&0.47 &   &Famaey et al. (2005)              \\
27.  & HD     20369\,AB   &--64.8 &6.9  & 1 &Fouts, Sandage (1986)             \\
    &                  &--65   &7    & 1 &Ryan, Norris (1991)               \\
32.  & HD     23439\,A    &+49.6 &1.8  & 3 &GCRV (1953)                       \\
    &                  &+50.92&0.09 &36 &Latham et al. (2002)               \\
    &                  &+50.70&0.10 & 1 &Nidever et al. (2002)              \\
    &                  &+49.8 &0.3  & 4 &Bartkevi\v cius, Sperauskas (1999)   \\
36.  & HD     26735\,A    &+64.3 &0.3  & 2 &Nordstrom et al. (2004)            \\
38.  & HD     27961\,AB   &+54.0 &5.0  & 1 &Nordstrom et al. (2004)            \\
50.  & HD     40412     &--46.5 &0.2  & 4 &Nordstrom et al. (2004)            \\
59.  & HD     60820\,AB   &+63.78&0.22 &   &Famaey et al. (2005)               \\
64.  & HD     74861\,AB   &--25.6 &0.2  & 3 &Nordstrom et al. (2004)            \\
66.  & HD     75632\,AB   &+47.5 &1.8  & 3 &GCRV (1953)                       \\
    &                  &+39.6 &0.3  & 2 &Wilson (1967)                     \\
    &                  &+43.5 &10.  & 1 &Reid  et al. (1995)                \\
    &                  &+44.9 &1.5  & 1 &Gizis et al. (2002)                \\
    &                  &+41.82&0.34 & 2 &Tokovinin, Smekhov (2002)         \\
88.  &PPM     11774     &--34.8 &0.4  & 2 &Bartkevi\v cius, Sperauskas (1999)   \\
 5V. & BD  +75  348\,AB   &+63.0 &     &   &Yamashita (1972)                  \\
    &                  & Orbit &  &  & Zacs et al. (2005)
\\
 7V. & HD    117466\,AB   & --1.39&0.23 &   &Famaey et al. (2005)               \\
    &                  & --2.0 &3.2  & 3 &Sperauskas, Bartkevi\v cius (2002)   \\
\tablerule
\end{tabular}
}
\end{center}
\vskip3mm

\begin{center}
\vbox{\footnotesize
\tabcolsep=2pt
\begin{tabular}{rllcccrcl}
\multicolumn{9}{c}{\parbox{110mm}{\baselineskip=8pt
{\normbf\ \ Table 5.}{\norm\ Basic data of radial velocity
variables.}}}\\
\tablerule
No. & Star name & CCDM & HIP & RA (2000) & ~DEC (2000) & $V$~~ &
$B$--$V$ & Sp \\
\tablerule
\noalign{\vskip2mm}
1. &TYC 2267-1300-1 &00046+3416\,D &      &00 04 33.5&+34 15 04&10.55  & 0.51  &           \\
2. &  BD  +69  230\,A &03566+6951\,A & 18448&03 56 36.2&+69 50 56& 9.33  & 0.93  &K0         \\
3. &  HD     29696  &04418+2840\,A & 21845&04 41 47.2&+28 39 36& 8.85  & 0.51  &F8         \\
4. &  BD  +39 1828\,AB&07036+3941\,AB& 34025&07 03 33.9&+39 40 33& 9.58  & 1.02  &M0         \\
5. &  BD  +75  348\,AB&            & 43042&08 46 11.6&+74 32 31& 9.55  & 1.12  &C3,0ch     \\
6. &  BD  +30 2129\,A &11171+2919\,A & 55115&11 17 03.5&+29 19 25& 9.86  & 0.57  &           \\
7. &  HD    117466\,AB&            & 65887&13 30 22.6&+07 24 54& 7.63  & 0.90  &K0         \\
8. &  BD  +82  565\,A &18471+8244\,A & 92162&18 47 02.6&+82 43 30& 9.34  & 0.66  &G0         \\
9. &  BD  +28 4035\,AB&21159+2858\,AB&104994&21 15 55.0&+28 57 47&10.41  & 0.81  &G5         \\
\tablerule
\end{tabular}
}
\end{center}

\noindent {\smallbf Notes}
\vskip3mm
{\footnotesize

\noindent 1. TYC 2267-1300-1, GSC 2267-1300, HIP 375D; AD: $PA$ = 238\degr, $d$ = 95.7\arcsec, $\Delta m$ = 1.7 mag

\noindent 2.   BD  +69  230A, G 247-8, LDS 1583,  SAO 12957, PPM 14567, LTT 11291; HIP-G; AB: $PA$ =

336\degr, $d$ = 22.1\arcsec, $\Delta m$ = 5.5 mag, Makarov \&
Kaplan (2005)            \\

\noindent 3.   HD     29696, HIP-C, $PA$ = 216.38\degr, $d$ = 28.84\arcsec, $\Delta Hp$ = 2.16 mag; B: HIP 21842, $V$ = 10.92

 mag

\noindent 4.   BD  +39 1828\,AB, HIP-C; Orbit: $P$ = 32.11 yr, $a$ = 0.247\arcsec, $\Delta V$ = 0.26 mag

\noindent 5.   BD  +75  348\,AB, SAO 6630

\noindent 6.   BD  +30 2129\,A, HIP-C, AB: $PA$ = 325.0\degr, $d$ = 3.960\arcsec, $\Delta Hp$ = 2.567 mag

\noindent 7.   HD    117466\,AB, BD +8 2720, SAO 119968, PPM 159812; HIP-G

\noindent 8.   BD  +82  565\,A, G\,259-37, LDS 1894; AB: $PA$ = 80\degr, $d$ = 12.0\arcsec, $\Delta m$ = 5.9 mag

\noindent 9.   BD  +28 4035\,AB, LTT 16237; HIP-C, AB: $PA$ = 86\degr, $d$ = 0.188\arcsec, $\Delta Hp$ = 1.93 mag
}
\vskip3mm

\begin{center}
\vbox{\footnotesize
\tabcolsep=2pt
\begin{tabular}{llrrrrrrr}
\multicolumn{9}{c}{\parbox{90mm}{\baselineskip=8pt
{\normbf\ \ Table 6.}{\norm\ Radial velocity variables.}}}\\
\tablerule
No. & Star name & ~~$<{V_r}>$ & $e$~~ & $E$~~ & $E/I$ & Span & $<\Delta
Vr>$ & $N$ \\
    &           &  km/s     & km/s & ~km/s &     &  d~~  & km/s~~  &  \\
\tablerule
\noalign{\vskip2mm}
 1. & TYC 2267-1300-1  &      --23.9 &    2.4  &   7.3  &   8.5& 1218& 18.7~~ &  9   \\
 2. &  BD  +69  230\,A   &       14.3 &    1.8  &   3.0  &   4.0&  107&  6.5~~ &  3            \\
 3. &  HD     29696    &        9.7 &    2.8  &   3.9  &   4.2&    5&  5.6~~ &  2            \\
 4a. &  BD  +39 1828\,A   &       33.4 &    0.2  &   1.0  &   1.1& 1068&  3.6~~ & 20            \\
 4b. &  BD  +39 1828\,AB  &       37.5 &    2.3  &   8.5  &  10.4& 1072& 33.7~~ & 14            \\
 4c. &  BD  +39 1828\,B   &       16.7 &    4.5  &  21.4  &  25.5& 1070& 62.6~~ & 23            \\
 5. &  BD  +75  348\,AB  &       50.5 &    1.8  &   3.2  &   4.4&  483&  6.1~~ &  3            \\
 6. &  BD  +30 2129\,A   &      --40.6 &    3.9  &  16.8  &  19.8& 1170& 49.8~~ & 19            \\
 7. &  HD    117466\,AB  &       --3.2 &    1.2  &   4.4  &   6.9& 1529& 10.4~~ & 13            \\
 8. &  BD  +82  565\,A   &      --79.0 &    2.4  &  18.7  &  24.1&  619& 58.6~~ & 60      \\
 9. &  BD  +28 4035\,AB  &      --47.9 &    3.1  &  11.1  &  12.0& 1184& 51.5~~ & 13            \\
\tablerule
\end{tabular}
}
\end{center}
\vskip3mm

Radial velocity variability for stars BD +69 230 and HD 29696, having
only three and two velocity observations respectively, will be verified
by supplementary observations.

For the ``CH-like" giant (C3,0ch) BD +75 348\,AB we obtained $<{V_r}>$ =
50.5$\pm$1.8 km/s, $E/I$ = 4.4, $<\Delta Vr>$ = 6.1 km/s, span =
483$^{\rm d}$, $n$ = 3. Recently Zacs et al.  (2005) presented
spectroscopic orbit for this star.

Stars HD 27961AB and HD 75632AB are included in the constant velocity
stars (Tables 1--3).  The first star has only one  observation
${V_r}$ = 36.3$\pm$0.7 km/s. A quite different value ${V_r}$ = +54.0
$\pm$ 5.0 km/s (also one observation) has been obtained by Nordstrom et
al.  (2004).  For HD 75632\,AB our two observations give $<{V_r}>$ =
46.7 $\pm$1.1 km/s, $E/I$ = 2 .4, $<\Delta Vr>$ = 2.2 km/s.  The old R.
E. Wilson's {\it General Catalogue of Stellar Radial Velocities} (GCRV)
gives the mean value from three authors $<{V_r}>$ = +47.5$\pm$1.8 km/s,
the result similar to ours.  However, Wilson (1967) in the
publication devoted to radial velocity measurements of dK and dM stars,
from two spectra gives a different value of $<{V_r}>$ = +39.6$\pm$0.3
km/s.  Other authors give intermediate values:  Reid et al.  (1995) give
+43.5$\pm$10 km/s from one measurement, Gizis et al.  (2002) give
+44.9$\pm$1.5 also from one measurement, and Tokovinin \& Smekhov (2002)
give the mean from two measurements $<{V_r}>$ = +41.82$\pm$0.34 km/s.
Consequently, for the confirmation of radial velocity variability, new
measurements are urgently needed for these thick disk orbital binary
stars.

A detailed discussion on physical or optical nature of the observed
double stars will be presented in other publications.  Here we limit our
discussion only by short notes about some observed binaries.

1. CCDM 00046+3416\,ABCD.  Component C (HIP 375) is optical.  Component
D is TYC 2267-1300-1.  A radial velocity variable.  From nine
observations ranging over 1218 days, the mean radial velocity is
$<{V_r}>$ = --23.9$\pm$2.4 km/s, quite close to the mean velocity of the
system AB (HD 225220\,AB), $<{V_r}>$ = --24.4$\pm$0.4 km/s, but for
definite conclusion radial velocity curve of component D is needed.

2. CCDM 00355+2431\,AB.  (BD +23 80\,AB).  A physical pair from radial
velocity observations.

3. CCDM 01066+6240\,AB (HD 6448\,AB). Physical system.

4. CCDM 02187+3429\,AB (HD 14202\,AB). Physical system.

5. CCDM 03470+4126\,ABC (HD 23439\,ABC).  Component C is clearly
optical.

6. CCDM 04157+4524\,AB (HD 26735\,AB).  A moderately high radial
velocity physical pair.

7. CCDM 04418+2840\,AB (A: HD 29696, HIP 21845; B: HIP 21842).
Component A is a possible radial velocity variable. More
observations are needed.

8. WDS 05329+5208\,ACB (AC: HD 36195). Physical triple.

9. WDS 05594+1749\,AB (A: HD 40412, B: HIP 120002). Physical double.

10. CCDM 19037+1658\,AB (A: HD 177349, B: HIP 93600). Optical pair.

11. CCDM 20356+2523\,AB (A: HDE 340730, B: GSC  2161-755). Possible
optical pair. More observations are needed.

\sectionb{6}{CONCLUSIONS}

Radial velocities for 91 mostly {\it Hipparcos} stars, the majority of
which are high transverse velocity doubles or multiples without previous
(at the moment of compilation of the observational program) or with low
precision or differing measurements, were obtained with a CORAVEL-type
radial velocity spectrometer.  The mean errors of the weighted mean
velocities are $<\varepsilon >$ = 0.60$\pm$0.02 km/s, $<{E/I}>$ =
0.79$\pm$0.07.  A comparison with the precise ($\varepsilon$$\le$ 2.7
km/s) 17 common radial velocities from the literature, does not show
zero point differences.  The discussion on ten detected radial velocity
variables is presented.  Component A of the visual binary BD +82 565 is
a thick-disk short-period ($P$ = 12.7 d) single-lined spectroscopic
binary.  A preliminary spectroscopic orbit ($P$ = 5.3 d) is determined
for component B of the visual binary BD +39 1828.  The components of the
four visual binaries HD 117466, BD +28 4035, BD +30 2129 and TYC
2267-1300 are found to be radial velocity variables and two more -- HD
29696 and BD +69 230 -- are possible radial velocity variables.  Our
measurements, as well as the data from the literature, show that the two
thick-disk orbital binaries, HD 27961\,AB and HD 75632\,AB, are also
radial velocity variables.

\thanks{J.  Sperauskas wishes to thank Tubitak National Observatory for
observing time on the RTT\,150 telescope and the Steward Observatory for
the use of the 1.5, 1.53 and 2.3 m telescopes.  J. Sperauskas is also
grateful
for the Vatican Observatory Research Group and the Jesuit Community at
Tucson for hospitality.  We are indebted to V-.D.  Bartkevi\v cien\.e
for preparation of the manuscript.  We acknowledge the use of facilities
of the Strasbourg Stellar Data Center (CDS), the NASA Bibliographic Data
Center (ADS), the Archive of the astrophysics preprints and the
Washington Visual Double Stars Catalog (WDS).}

\newpage

\References

\refb Barbier-Brossat M., Figon P. 2000, {\it General Catalog of
Averaged Stellar Radial Velocities for Galactic Stars}, A\&AS, 142, 217;
CDS Catalog No. III/213

\refb Bartkevi\v cius A. 1980, {\it The Catalogue of Metal-deficient
F-M Stars.  Part I. Stars Classified Spectroscopically (MDSP)}, Bull.
Vilnius Obs., No. 51, 3

\refb Bartkevi\v cius A. 1984, {\it Catalogue of Metal-deficient F--M
Stars. Part I. Stars Classified Spectroscopically.  Supplement 1
(MDSPS1)}, Bull.  Vilnius Obs., No. 68, 3

\refb Bartkevi\v cius A. 1992, {\it Catalogue of Metal-deficient F--M
Stars. II. Stars Classified Photometrically (MDPH Catalogue)},  Baltic
Astronomy, 2, 294

\refb Bartkevi\v cius A. 1994, Baltic Astronomy, 3, 34

\refb Bartkevi\v cius A. 2000, {\it The Bibliographic Catalogue of
Stellar Radial Velocities for Population II and Late-type Stars in the
Galaxy}, preprint.  Institute of Theoretical Physics and Astronomy,
Vilnius

\refb Bartkevi\v cius A., Bartkevi\v cien\.e D. 1993, {\it A Preliminary
Version of the Catalogue of Field Population II Stars on Magnetic Tape
(POP2 Catalogue)}, Bull.  Inform.  CDS, No. 42, 19--26; CDS, Strasbourg,
Magnetic Tape S5072

\refb Bartkevi\v cius A., Gudas A. 2001, Baltic Astronomy, 10, 481

\refb Bartkevi\v cius A., Gudas A. 2002, Baltic Astronomy, 11, 153

\refb Bartkevi\v cius A., Sperauskas J. 1990, Proc.  Nordic-Baltic
Astron.  Meeting, eds.  C.-I.  Lagerkvist, D. Kiselman \& M. Lindgren,
Uppsala, p. 45--48

\refb Bartkevi\v cius A., Sperauskas J. 1994, Baltic Astronomy, 3, 49

\refb Bartkevi\v cius A., Sperauskas J. 1999, Baltic Astronomy. 8, 325

\refb Bartkevi\v cius A., Sperauskas J. 2003, Astron.  Nachr., 324, 460
(BS 2003)

\refb Bartkevi\v cius A., Sperauskas J., Rastorguev A. S., Tokovinin
A. A. 1992, Baltic Astronomy, 1, 47

\refb Bidelman W.P. 1998, PASP, 110, 268

\refb Chiu L.-T. G. 1980, ApJS, 44, 31

\refb Duflot M., Fehrenbach C.,  Mannone C., Burnage R., Genty V. 1992,
 A\&AS, 94, 479

\refb ESA. 1997.  {\it The Hipparcos and Tycho Catalogues.  Double and
Multiple Star Annex}

\refb Famaey B., Jorissen A., Luri X., Mayor M., Udry S., Dejonghe H.,
Turon C. 2005, A\&A, 430, 165

\refb Fouts G., Sandage A. 1986, AJ, 91, 1189

\refb Gizis J. E., Reid I. N., Hawley S. L. 2002, AJ, 123, 3356;  CDS
Catalog,  J/AJ/123/3356

\refb Griffin R. F. 1967, ApJ, 148, 465

\refb Jasniewicz  G.,  Mayor M. 1988, A\&A, 203, 329

\refb Jones B. F. 1972, ApJ, 173, 671

\refb Latham D. W., Stefanik R. P., Torres G., Davis R. J., Mazeh T.,
Carney B. W., Laird J. B., Morse J. A. 2002, AJ, 124, 1144

\refb Loth A. L., Bidelman W. P. 1998, PASP, 110, 268

\refb Luyten W. J. 1922, Bull Lick Obs., No. 336

\refb Makarov V. V., Kaplan G. H. 2005, AJ, 129, 2420

\refb Malaroda S., Levato H., Morrell N., Garcia B., Grosso M.,
Bolzicco G. J. 2000.  {\it Bibliographic Catalogue of Stellar Radial
Velocities (1991--1994)},  A\&AS, 144, 1; CDS Catalog, No. III/204

\refb Nidever D. L., Marcy G. W., Butler R. P., Fischer D. A., Vogt S.
S. 2002, ApJS, 141, 503

\refb Nordstrom B., Mayor M., Andersen J., Holmberg J., Pont F.,
Jorgensen B. R., Olsen E. H., Udry S., Mowlavi N. 2004, A\&A, 419, 98

\refb Reid I. N., Hawley S. L., Gizis J. E. 1995, AJ, 110, 1838;  CDS
Catalog,  III/198

\refb Ryan S. G., Norris J. E. 1991, AJ, 101, 1835

\refb Sperauskas J., Bartkevi\v cius A. 2002, Astron. Nachr., 323, 139
(Paper I)

\refb Sperauskas J., Bartkevi\v cius A. 2005, Baltic Astronomy, 14, 527
(this issue)

\refb Tokovinin A. A. 1987, AZh, 64, 196

\refb Tokovinin A. A., Duquennoy A., Halbwachs J.-L., Mayor M. 1994,
A\&A, 282, 831

\refb Tokovinin A. A., Smekhov M. G. 2002, A\&A, 382, 118; CDS Catalog,
J/A+A/382/118

\refb Udry S., Mayor M., Maurice E., Andersen J., Imbert M., Lindgren
H., Mermilliod J.-C. 1999, in {\it Precise Stellar Radial Velocities},
eds.  J. B. Hearnshaw \& C. D. Scarfe, ASP Conf.  Ser., 185, 383 (IAU
Colloq.  No. 170)

\refb Upgren A., Sperauskas J., Boyle R. P. 2002, Baltic Astronomy, 11,
1

\refb Wilson R. E. 1953, {\it General Catalogue of Stellar Radial
Velocities  (GCRV)},  Carnnegie Institution Publ.,  Washington.

\refb Wilson O. C. 1967, AJ, 72, 905

\refb Woolley R., Penston M. J., Harding C. A., Martin W. L., Sinclair
J. E., Haslan C. M., Aslan S., Savage A., Aly K., Asaad A. S. 1981,
Royal Obs. Annals, No. 14

\refb Yamashita Y. 1972, Ann. Tokyo Obs., 13, 169

\refb Zacs L., Schmidt M. R., Glazutdinov G. A., Sperauskas J. 2005,
A\&A, 441, 303

\end{document}